\documentstyle[11pt,newpasp,epsf]{article}
\begin{document}
\title{The enigma of lithium: from CP stars to K giants. First results of 
CP star observations obtained at Mount Stromlo Observatory}

\author{N.S. Polosukhina\altaffilmark{1}, N.A. Drake\altaffilmark{2,3}, 
M. Hack\altaffilmark{4}, R. de la Reza\altaffilmark{2},}
\author{P.R. Wood\altaffilmark{5},
A.V. Shavrina\altaffilmark{6}}

\affil{$^1$Crimean Astrophysical Observatory, Nauchny, Crimea, 334413, Ukraine}
\affil{$^2$Observat\'orio Nacional/MCT, Rua General Jos\'e Cristino 77,
    20921-400, Rio de Janeiro, Brazil}  
\affil{$^3$Sobolev Astronomical Institute, St.\,Petersburg State University,
        Universitetsky pr. 28, Petrodvorets, 198504, St. Petersburg, Russia}
\affil{$^4$Department of Astronomy, Trieste University, Via Tiepolo 11, 34131 
       Trieste, Italy} 
\affil{$^5$Research School of Astronomy and Astrophysics, Mount Stromlo
      Observatory, Australian National University, Weston Creek, ACT 2611, Australia}
\affil{$^6$Main Astronomical Observatory of National Academy of Sciences,
          Kiev 252650, Ukraine}

\begin{abstract}

We present the results of the observations of some roAp stars
made at Mount Stromlo Observatory  during 17 nights
in 2001 September-October. This long observing run permitted us to
obtain a good phase-rotation coverage.
In chemically peculiar magnetic stars, the Li\,{\sc i} 6708~\AA\ spectral
line presents very anomalous behaviour: 
in some stars it is a strong feature, in others, with
similar atmospheric parameters, it is invisible.
Interesting results were obtained for the roAp star HD\,3980 which
presents variations of the profile and position of the Li\,{\sc i} line
with the rotation period.
These new observational results should serve as a base for
the development of atmospheric models of ``Li-spotted'' roAp stars.

\end{abstract}


\index{*HD 3980|ksi Phe}
\index{*HD 15144|AB Cet}
\index{*HD 24712|DO Eri}
\index{*HD 83368|IM Vel}
\index{*HD 42659|UV Lep}
\index{*HD 208217|BD Ind}
\index{*HD 201601|gamma Equ}

\section{Introduction}

The ``Li puzzle'' is the great spread in lithium abundance for stars with similar
physical parameters
($T_{\rm eff},\; \log g, \;M$).  Its main cause is related to unknown physical processes.
The strongest Li feature in a star's spectrum -- the lithium 
resonance doublet at 6708 \AA\ -- is very sensitive to evolutional changes,
to the temperature regime and to conditions of mixing.
Usually, lithium is depleted with stellar age. The presence of the Li\,{\sc i} line 6708 \AA\ in a stellar
spectrum is an indication of youth of a star, or of a breaking of mixing between 
internal (hot) and external (cool) layers of a star, or an indication of active
processes with an eventual lithium synthesis.

The presence of a magnetic field is one of the conditions for Li synthesis (for example, by spallation
reactions on the stellar surface). A magnetic field also inhibits mixing of stellar material and,
hence, lithium depletion.

An influence of surface activity connected with magnetic field structure on
Li line profiles is a problem under discussion for the late type chromospherically
active Li-rich giants too. 
Attempts to detect spots and rotational modulation with photometric 
variations gave  contradictory results (Randich et al. 1993).
Since the discovery of the first Li-rich K giant, the existence of oxygen giant stars 
with high Li abundance became a puzzle, although different mechanisms were proposed 
to explain the high Li abundance on the surface of these stars (Brown et al. 1989; de la Reza et al. 1996, 1997).
It is important to  mention that in the past
Lambert \&  Sawyer (1984) suggested that Li-rich giants may be the ``descendants'' of one or
more classes of magnetically peculiar CP stars, i.e. there is an evolutional connection
between magnetic CP stars with high Li abundance and Li-rich red giants.

A comprehensive investigation of the correlation of lithium line profiles with chemically
peculiar (CP) stars' rotation (Polosukhina et al. 1999) permits us to separate 
the observed Ap-CP stars into four groups with different behaviour of the Li\,{\sc i} line 6708 \AA.
The behaviour of this line can be explained by the existence of Li-rich spots on the star's
surface, using the oblique rotator model with different parameters for each star. 
Recently, a preliminary spectroscopic analysis of the chemically peculiar Ap star HD\,83368 
was carried out in the Li\,{\sc i} 6708 \AA\ spectral region (Polosukhina et al. 2000).
This analysis revealed a very strong Li line 6708 \AA\ with significant 
variations in the intensity and position with rotational phases. 
The Li line was attributed to two lithium spots (with $\log \epsilon ({\rm Li})$ 
equal to 3.6 and 3.5, respectively).
The Li spottedness on the surface of HD\,83368 was first indicated
 by North et al. (1998).

The lack of recent advances in the interpretation of the Li behavior in peculiar Ap 
stars could be ascribed to the scanty number of available observations.
The recent very important and interesting observations of CP stars 
at ESO gave a new impulse for the organization of a new monitoring campaign of CP 
stars in the Li spectral region.

\section {Observations}

High resolution ($R=47\,000$) spectra of HD\,83368
were obtained with FEROS at 1.52m telescope in ESO, Chile.
Observations were carried out on December 5 to 12, 2000. Due to the short rotational
period of the HD\,83368 ($P=2.\!\!^d852$), two  phases were
observed every night. Spectral coverage is  3900 \AA\ -- 9000 \AA\,
without gaps. At the present time the spectral regions of the 
Li\,{\sc i} line at 6708 \AA\ and O\,{\sc i} triplet at 7770 \AA\ have been analyzed
(see Kochukhov, Drake, de la Reza in this Proceedings). 

High-resolution spectra ($R=88\,000$) of eleven CP stars were obtained during  
the observing run in  September -- October 2001
 with the 74-inch telescope and echelle spectrograph of the Mount Stromlo Observatory. 
Now we present the first part of the research concentrated on the
Li\,{\sc i} 6708 \AA\ feature.

The reduction of the observed spectra from bias substraction and 
flat-field corrections through spectra extraction and wavelength calibration
was done using standard NOAO IRAF procedures.

\section {\bf Results}

Analysis of the observational results permits us to separate
our sample into two groups: stars with a clearly detectable Li\,{\sc i}
6708 \AA\ line and stars having no or a very weak feature at 6708 \AA.
Some stars from our program, such as HD\,3980, HD\,83368, HD\,208217, 
and $\gamma$ Equ, show a clearly detectable feature at 
the predicted position of the Li\,{\sc i} doublet. 
However, the spectra of some stars have not yet been
reduced.

In order to identify the nature of the 6708 \AA\ feature, we overlapped 
the observed spectra and synthetic ones calculated with the atmospheric 
parameters of a given star
(previously determined by us or taken from the literature).
Atmospheric models, atomic data, and synthetic spectra computing code are those as in
Polosukhina et al. (2000). 
For many lines present in the spectra of CP stars, we do not know atomic parameters
or even their identifications.  So, some spectral lines were not included in the
calculations of the synthetic spectra. 
The lines of the rare earth elements (REE) are very strong in the spectra
of CP stars, and they were included in our line list.
For example, the strong Ca\,{\sc i} line at 6717.681 \AA\ is blended with 
Gd\,{\sc ii} 6718.13 \AA. This line is very strong 
in the spectra of Ap -- CP stars, such as HD\,3980, HD\,15144, and HD\,208217,
while in the spectra of normal A - F stars the line at 6718 \AA\ is due to the
Ca\,{\sc i} 6717.681 \AA\ 
line blended with a weak Ti\,{\sc i} 6717.794 \AA\ line.

The line of Pr\,{\sc iii} 6706.705 \AA\ near the Li feature is clearly
visible in the spectra of some roAp -- CP stars and is a strong indicator 
of roAp stars.
Some stars, such as $\gamma$ Equ and HD\,24712, were observed by us for 
testing and comparison with the results of other observers.

Here we present only some part of observations and preliminary results 
of the spectral reduction. The final analysis of all obtained results will 
be presented elsewhere.

Now we present only the spectra of the Li\,{\sc i} 6708 \AA\ line region. 
In the future we plan to study other spectral regions containing  lines of
some alkaline  elements having excitation conditions similar to those of the Li line. 
An analysis of temperature sensitive lines may permit us to separate 
the effects of the temperature and abundance inhomogeneities on the line 
profiles  of the studied elements.

In Figures 1 -- 7 we present original spectra of some CP stars in the 
Li\,{\sc i} 6708 \AA\ spectral region.

\begin{figure}
\plotone{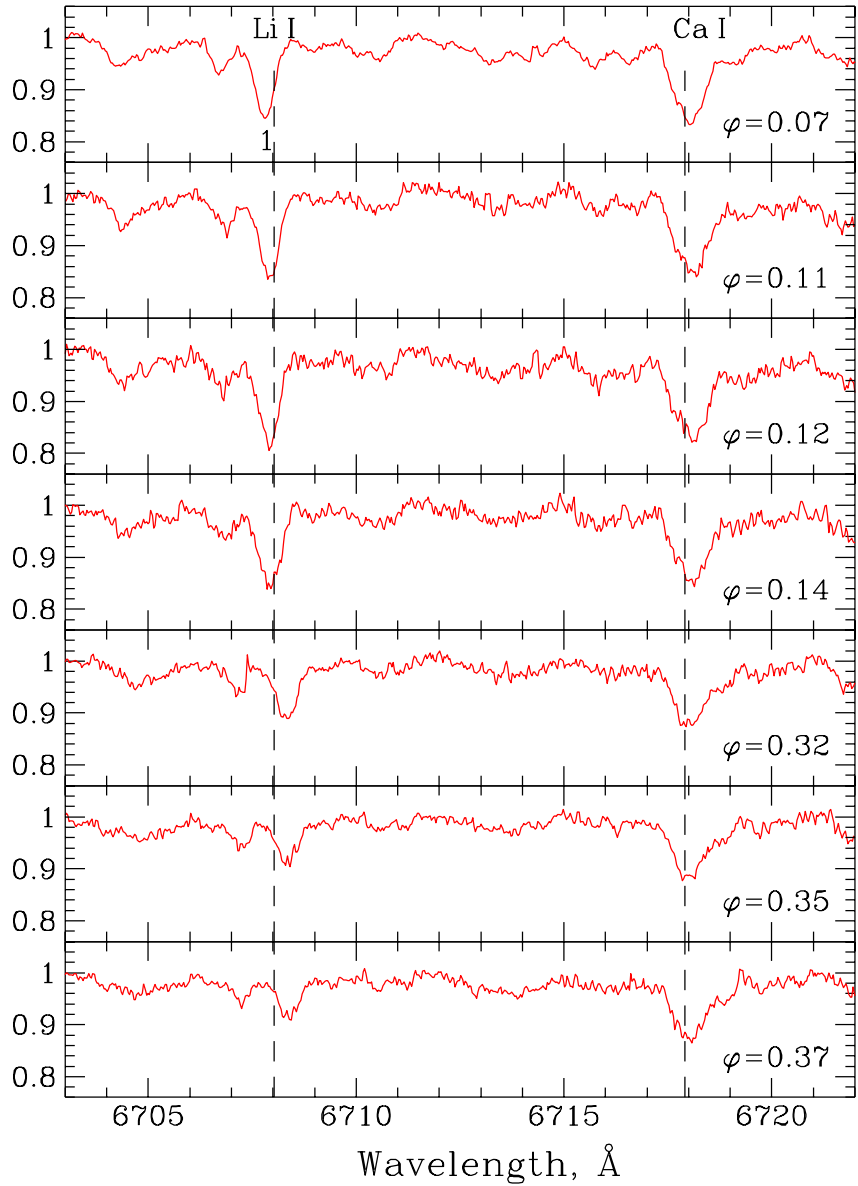}
\caption{Spectra of the star HD\,3980 in the Li\,{\sc i} 6708 \AA\
region. The rotational phases are given on the right. 
The Li\,{\sc i} line shows strong variability both in position  and in intensity.
In our interpretation, the Li line originated in a first ``Li spot'' (spot 1)
moves to the red from phase 0.07 to 0.37. A second ``Li spot'' (spot 2)
has appeared by phase 0.58 (see Fig. 2) and the Li line position moves to the red also.
Note the striking similarity of behaviour of Li line 6708 \AA\ profiles 
of HD\,3980 with Li-spotted star HD\,83368.}
\label{fig-1}
\end{figure}

\begin{figure}
\plotone{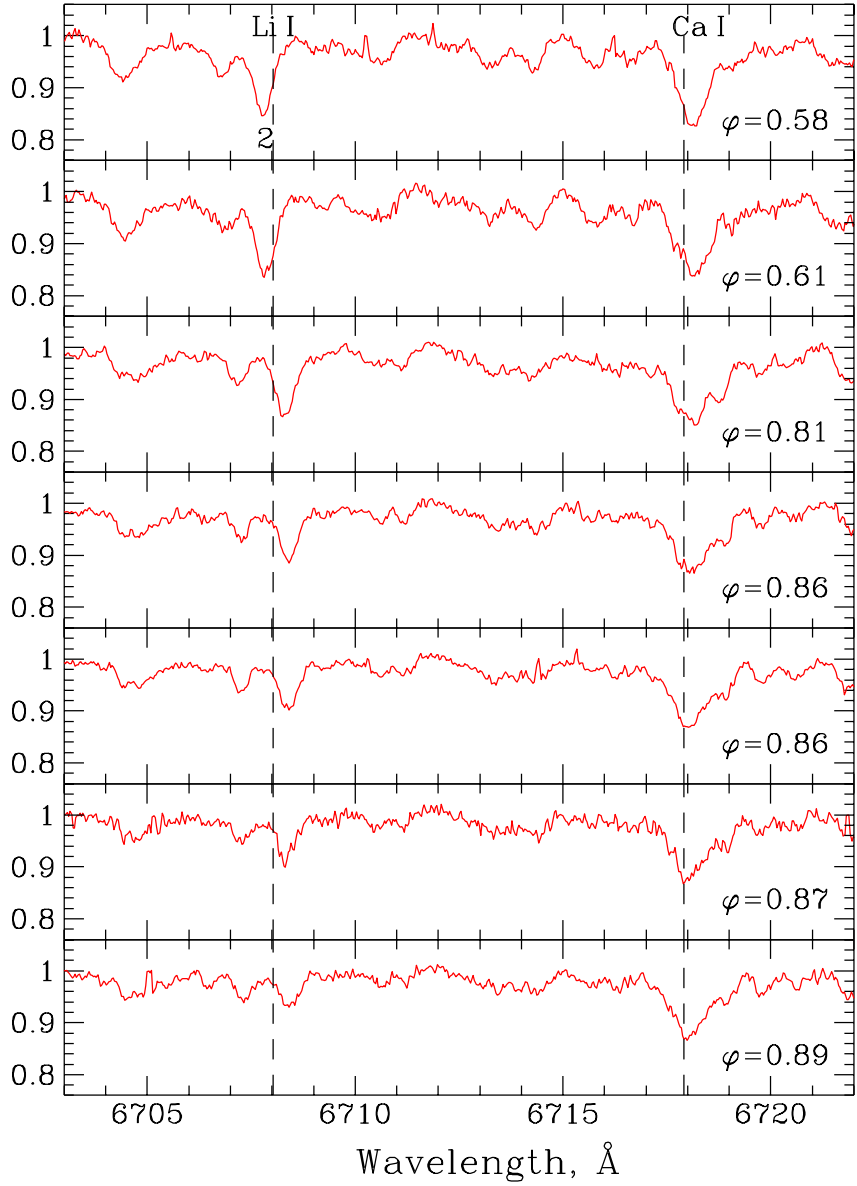}
\caption{Continuation of the spectra in Figure 1.}
\label{fig-2}
\end{figure}

\begin{figure}
\plotone{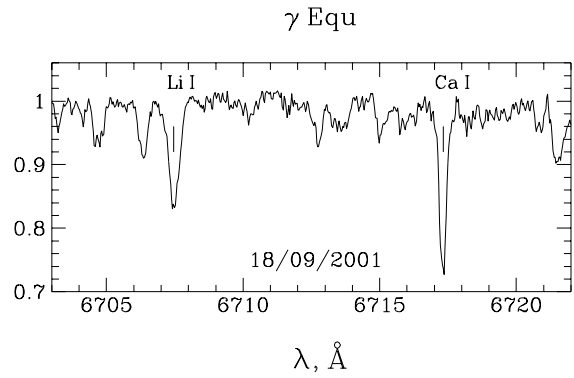}
\caption{The spectrum of the well known Li CP star $\gamma$ Equ.
We present this star for comparison reasons.}
\label{fig-3}
\end{figure}

\begin{figure}
\plotone{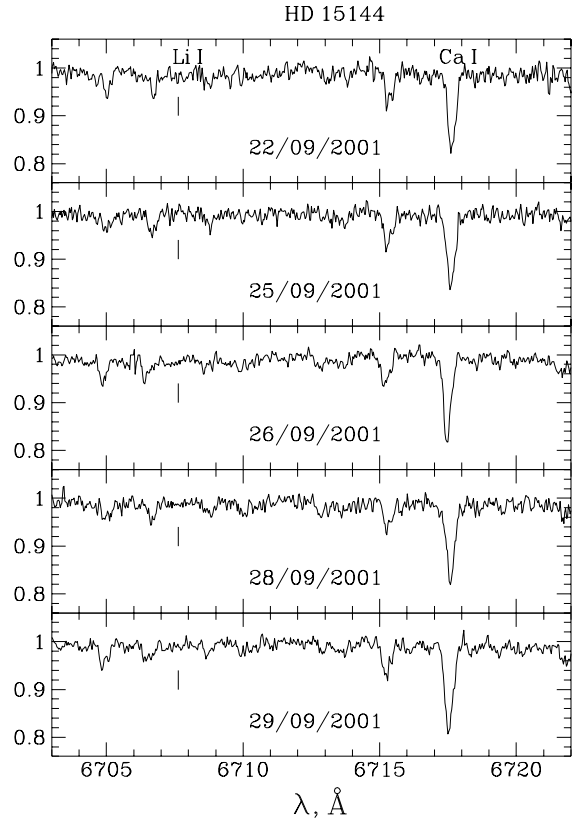}
\caption{Spectra of the star HD\,15144 in the Li\,{\sc i} 6708 \AA\
region. No Li line is observed.}
\label{fig-4}
\end{figure}

\begin{figure}
\plotone{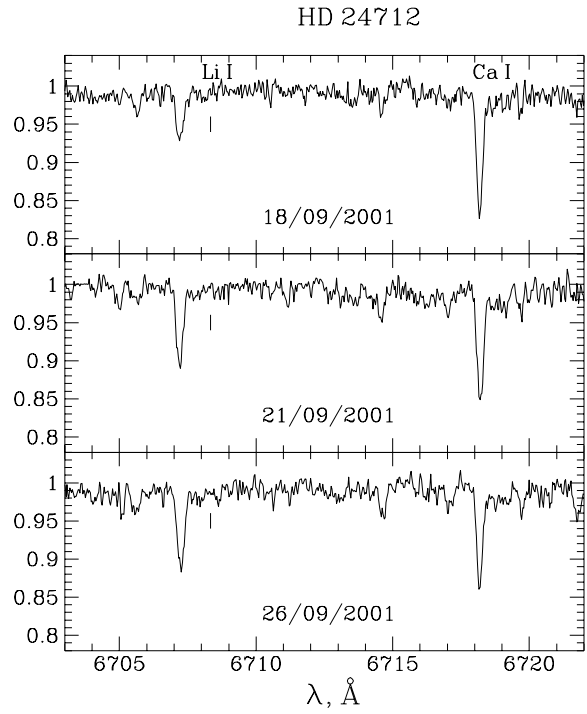}
\caption{Three spectra of the star HD\,24712.
Li\,{\sc i} line is not observed in this star, its position 
is marked by short vertical line.
A feature at $\sim 6707$ \AA\ is the Pr\,{\sc iii} line
6706.705 \AA.}
\label{fig-5}
\end{figure}

\begin{figure}
\plotone{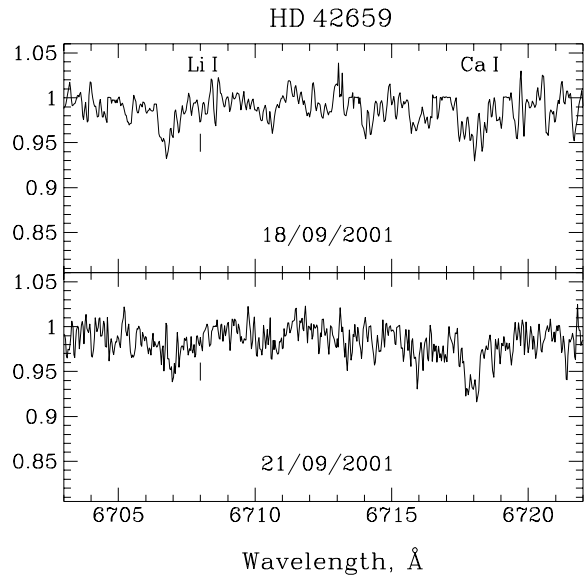} 
\caption{Two spectra of the star HD\,42659 obtained with  
time difference of three days. No Li\,{\sc i} line is observed.} 
\label{fig-6}
\end{figure}

\begin{figure} 
\plotone{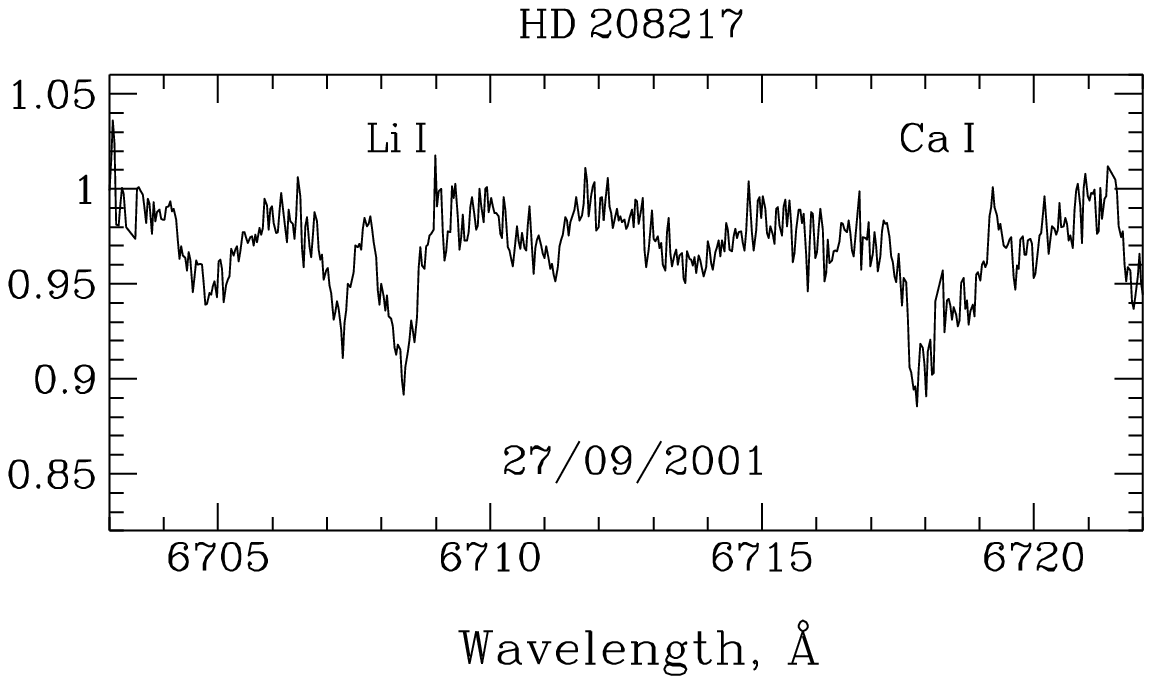} 
\caption{ HD\,208217. The strong Li\,{\sc i} line is observed 
in the spectrum of this star obtained on September 27.} 
\label{fig-7} 
\end{figure}

\section {Comments on Individual Stars} 

The most important result was obtained for the star HD\,3980. 
The monitoring of this star, as can be seen on Figures 1 and 2,  
shows strong variations of the spectra, especially in the Li\,{\sc i} 
6708 \AA\ line. 
A large Doppler shift of the Li line position is observed,  
as it was for HD\,83368. This gives us an opportunity to 
join these two stars in one group of CP stars with lithium spots. 
 
{\bf HD\,3980 = $\xi$ Phe} is a late type Ap star with strong 
photometric variations in $V$-band. On this curve 
there are two minima (primary minimum amounts  to 0.13 mag, the 
secondary minimum  shows half of this value). 
The two maxima are equal, and all extrema are separated by 0.25 
phases. We have not completed spectral studies concerning this star, 
but in the similar star, 
HD\,98088,  the lines of REE (for example, Eu\,{\sc ii} lines) vary 
exactly in antiphase with the 
light curve: primary and secondary minima of light variations correspond to primary  
and secondary maxima in the line variations. Both stars belong to A3 spectral type. 
In the paper of Maitzen et al. (1980) there are magnetic measurements of $H_{\rm eff}$  
(for HD\,3980) with a great scattering.  
A reversal of polarity was noted every second day, while $H_{\rm eff}$ 
is variable with a 4 day period and amplitude of 2 kG. Unfortunately, magnetic field  
measurements are not reliable enough to make a $H_{\rm eff}$ magnetic field curve.

We used for phase calculations the rotational period of $P=3.95200$ days 
(Catalano \&  Renson  1998). 
The radial velocity of this star, $V_{\rm rad} =9.8\;{\rm km\,s^{-1}}$, 
was taken from the catalog of Barbier-Brossat  \&  Figon  (2000).


{\bf HD\,83368 = IM Vel} is a bright, southern, rapidly oscillating 
(roAp), cool, magnetic,  chemically peculiar star. 
In addition to the short-term pulsational changes, the star shows variations of  
spectral line profiles, mean longitudinal magnetic field and brightness with a 
rotational period of 2.851976 days.
  
{\bf HD\,208217 = BD Ind} is a variable star of $\alpha^2$ CVn type having  
$P=8.44500$ days and  $v\sin i = 55.0\;{\rm km\,s^{-1}}$ (Catalano \&  Renson  1998). 
 At first, we detected a strong Li\,{\sc i} line in the  
spectrum of this star  obtained on September 27, 2001.  
We have a second spectrum of this star obtained two days before, but, unfortunately,  
it is of very poor quality due to bad weather conditions. 
However, there are indications of strong  
variability of its spectrum. More observations are clearly needed.

{\bf HD\,201601 = $\gamma$ Equ} is the well known CP star of spectral type F0p. 
We present this star for comparison reasons. 
 
In the second group of CP stars, those with an absent  Li line, are the 
stars: HD\,15144, HD\,42659, and HD\,24712.  
Position of the Li\,{\sc i} line 
is marked by a short vertical line in the figures displaying their spectra.

\section {Discussion}

- During  our observations with the 74-inch telescope of the Mount Stromlo observatory
using the echelle spectrograph, we discovered that HD\,208217 is a CP star 
with strong Li\,{\sc i} line 6708 \AA. 
 
\smallskip 
 
- Reduction of all 74-inch spectra is unfinished, but the most important result 
of this set is a monitoring of HD\,3980 which is shown in Figures 1 and 2.  
 
\smallskip
 
- The behaviour of the Li\,{\sc i} 6708 \AA\ line in this star is very  
similar to that of HD\,83368.  
It is a new evidence of Li spottedness on CP star surfaces. We have  
finished only preliminary reduction of HD\,3980 data 
and obtained only an estimation of Li abundance $\log\epsilon({\rm Li}) \sim 4.0$. 
 
\smallskip 

- There are some stars with similar physical atmospheric parameters   
($T_{\rm eff}$,  $\log g$) but without the Li\,{\sc i} line at 6708 \AA:  
HD\,15144, HD\,24712, and HD\,42659. For these stars we carried out a monitoring too. 

\smallskip 
 
- Our knowledge of some peculiarities and magnetic field structure of 
these stars is very poor.
But there are some indications that HD\,3980 has a dipole magnetic field 
like HD\,83368. 

\smallskip 
 
- At the present time we have some CP stars with high Li abundance  
($3.1 \le \log\epsilon({\rm Li}) \le 4.0$): 
$\beta$ CrB, $\gamma$ Equ, HD\,188041, HD\,83368, HD\,60435,  HD\,134214,  
HD\,137949,  HD\,166413, HD\,101065, HD\,3980. 
Some of these stars were studied by Faraggiana, Gerbaldi,  \&  Delmas (1996). 
Our results for the Li abundance of HD\,188041, HD137949, and HD\,3980 are 
in good agreement with the results of these authors.
 
\section{Conclusions} 
 
The discovery of Li spots in HD\,83368 and HD\,3980 
is the first indication of spottedness in the lithium distribution 
in some cool magnetic CP stars. A good correlation between 
positions of the spots, magnetic field, brightness and oscillation 
phenomena indicates possible connection between the 
magnetic field configuration, the local structure of the atmosphere 
and the local distribution of the chemical anomalies, such 
as Li abundance. However, the present state of our knowledge does not allow us 
to make any more detailed conclusions about the structure 
and physical conditions in the atmosphere within or outside the 
magnetic spots. Model atmospheres which take into account  
magnetic fields, radiative and convective energy transport, opacity 
due to overabundances, etc., are needed. Also, the occurrence
of lithium on the surface needs to be studied. A theoretical possibility 
has been suggested by Babel (1994): ambipolar diffusion 
of hydrogen may result in a significant overabundance of Li\,{\sc i} in 
the vicinity of the magnetic poles of CP2 stars.

The investigation of the physical conditions in Li spots 
should be the next step in the study of the cool magnetic CP 
stars' anomalies. Lithium might be the ``key'' element to improve 
our understanding of the atmospheric structure and other 
anomalies in these stars.

\end{document}